\documentclass[twocolumn,bibnotes,amsmath,amssymb,aps]{revtex4-2}

\usepackage{graphicx}% Include figure files
\usepackage{bm}% bold math
\usepackage[per-mode=symbol]{siunitx}
\usepackage{physics}
\usepackage{braket}
\usepackage{relsize}
\usepackage{hyperref}% add hypertext capabilities
\usepackage{dcolumn}% Align table columns on decimal point
\usepackage{dsfont}
\newcommand{\ahat}[1]{\hat{a}_{#1}}
\newcommand{\adag}[1]{\hat{a}^\dag_{#1}}
\newcommand{\upa}{\uparrow}
\newcommand{\dna}{\downarrow}
\newcommand{\vc}[1]{{\vec{#1}}} %{{\bm{#1}}}
\newcommand{\br}{{\bm{r}}} 

\begin{document}

\title{Spin Rotations in a Bose-Einstein Condensate Driven by Counterflow and Spin-independent Interactions}
\author{David C.\ Spierings}
\altaffiliation[Currently at ]{MIT, Cambridge, MA}
%\email{dspierin@mit.edu}
\author{Joseph H.\ Thywissen}
%\email{joseph.thywissen@utoronto.ca}
\author{Aephraim M.\ Steinberg}%
%\altaffiliation[Also at ]{CIFAR, Toronto, Ontario, Canada.}
\affiliation{Department of Physics and CQIQC, University of Toronto, Toronto, Ontario M5S 1A7, Canada}

\date{\today}

\begin{abstract} 
We observe spin rotations caused by atomic collisions in a non-equilibrium Bose-condensed gas of $^{87}$Rb. Reflection from a pseudomagnetic barrier creates counterflow in which forward- and backward-propagating matter waves have partly transverse spin directions. Even though inter-atomic interaction strengths are state-independent, the indistinguishability of parallel spins leads to spin dynamics. A local magnetodynamic model, which captures the salient features of the observed spin textures, highlights an essential connection between four-wave mixing and collisional spin rotation. The observed phenomenon has previously been thought to exist only in nondegenerate gases; our observations and model clarify the nature of these effective-magnetic spin rotations. 
\end{abstract}

\maketitle

Spin dynamics in cold atomic gases exhibit rich phenomena due to the interplay of particle interactions, quantum coherence, and particle statistics. In a Bose-Einstein condensate (BEC) with only contact interactions, spin dynamics can be induced by the spin-dependence of the interaction strengths between particles. 
If inter-spin and intra-spin interaction strengths are the same, on the other hand, there is only one scattering length, $a$. When all atoms have the same kinetic energy and experience the same trapping potential regardless of spin, one might be tempted to conclude that all spin components have the same energy and therefore distribute themselves evenly following the density profile.

Wave-like spin excitations (i.e.~spin waves), however, have long been known to arise in Fermi gases and {\it nondegenerate} Bose gases even in the absence of spin-dependent interactions \cite{LR:1968,*Leggett:1970,bashkin1981,lhuillier1982I,lhuillier1982II}, and have been observed in a variety of systems \cite{johnson1984,nacher1984,gully1984,Lewandowski2002,McGuirk2002,Du:2008de,*Du:2009hm,Piechon:2009,*Natu:2009gt,Kohl2013,Trotzky2015}. Spin waves in dilute quantum gases are generated by the Identical-Spin Rotation Effect (ISRE), an effective magnetic interaction in which colliding spins precess about their net spin \cite{oktel2002,nikuni2002,Piechon:2009,*Natu:2009gt}. Both ISRE and the closely related Leggett-Rice effect~\cite{Miyake:1985hz} 
arise from the interaction energy difference when two identical particles collide with their spins aligned or anti-aligned due to exchange-symmetry contributions. Particle exchange modifies the interaction energy of a gas by $g\rho$, for particle density $\rho$ and spin-independent interaction strength $g=4\pi\hbar^2a/m$. 
In contrast, all particles in a BEC at zero temperature occupy the same single-particle state and so the many-body wavefunction can be written as one already-symmetrized product of single-particle wavefunctions (i.e.~Hartree form), devoid of exchange contributions or bunching \cite{oktel2002BEC,nikuni2003}. Thus, the conventional picture of spin rotation between identical particles would suggest that they should not occur in a condensate. 

In this work, we observe spin dynamics emerging in a BEC that is reflected from a pseudomagnetic barrier. Counter-propagating matter waves generate a density modulation that leads to spin-dependent four-wave mixing, which we show can also be recast as an effective-magnetic spin rotation, whether or not the system is degenerate. 
This equivalence arises because a two-body interaction term 
$\adag{+k,\dna} \adag{-k,\upa} \ahat{-k,\dna} \ahat{+k,\upa}$
can be seen as either the exchange of momentum between two spins or the exchange of spin between two momentum modes. 

Here we follow the latter picture and derive a magnetic interaction when position is coarse-grained over a length scale much longer than that of the density modulation $\pi/k$, such that the local density can be approximated as constant. In this case, the populations of momentum components $+k$ and $-k$ are conserved locally. Remarkably, it follows \cite{supp} that pseudospin-half bosons with two such momenta experience an effective magnetic interaction given by 
\begin{equation}
\begin{aligned} \label{eq:magneto}
\dot{\vc{\sigma}}_{+ k}(\br)&=g\rho_{- k}(\br)\, \vc{\sigma}_{- k}(\br)\bm\times\vc{\sigma}_{+ k}(\br) \\
\dot{\vc{\sigma}}_{- k}(\br)&=g\rho_{+ k}(\br)\, \vc{\sigma}_{+ k}(\br)\bm\times\vc{\sigma}_{-k}(\br) 
\end{aligned}
\end{equation}
where $\vc{\sigma}_{\pm k}(\br)$ are position-dependent Bloch vectors for the $\pm k$ modes, and $\rho_{\pm k}(\br)$ are the average densities of each momentum mode, summed over spin states. In this local magnetodynamic (LMD) model, the Bloch vector of each momentum mode  rotates about the Bloch vector of the other, with a precession frequency determined by the counter-propagating mean field, $g\rho$.

In condensates with counterflow, the bosonic bunching observed in thermal systems is not present. Nonetheless, for identical spins, there are density fluctuations due to interference. In both cases, the particle-particle correlations lead to increased  interaction energy. 
Unlike ISRE, the spin-dependent dynamics arising from four-wave mixing persist into the mean-field limit and can be captured by a Gross-Pitaevskii (GP) treatment. 
Interfering particles see a higher average density because they spend more time near the peaks of the density distribution. By comparison, the orthogonally polarized component of the counterflow has a spatially uniform probability density. 
Hence, the interaction energy is different for the two spin states. This difference can be on the order of the energy of an entire condensate, since it includes contributions from all pairs of atoms, just as in the case of fragmentation \cite{nozieres1982,Mueller2006}. 
The indistinguishability of particles permits both interference and exchange effects like ISRE. 

Spatially inhomogeneous dynamics of two-component BECs has been studied in systems where component separation is created via the trapping potential \cite{hall1998,cornell1998having} or induced by differences in interaction strengths \cite{Mertes2007,Anderson2009}. Furthermore, component separation has been used as a mechanism for spin-squeezing via the one-axis twisting Hamiltonian for systems with nearly equal interaction strengths \cite{riedel2010,schmied2016,haine2014,Laudat2018}. Yet, the effective magnetic interaction typically ascribed to ISRE has not been directly observed in a condensed system, even though self-rephasing, a byproduct of this interaction, has \cite{Deutsch2010,Egorov2011}. 

\begin{figure}[t]
\includegraphics[width=\columnwidth]{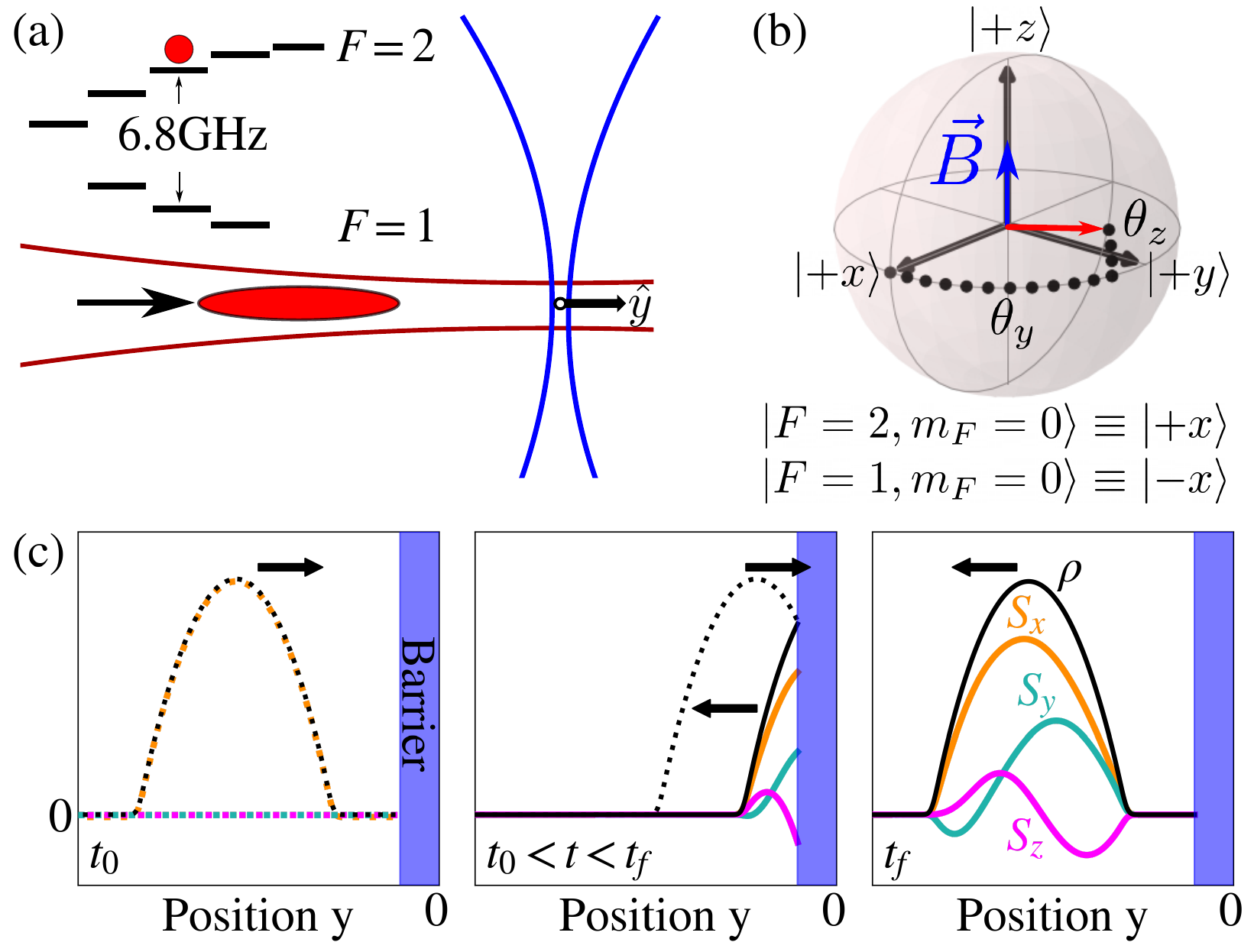}
\caption{{\bf Experimental configuration.} \label{fig:setup} 
(a) Illustration of the quasi-1D scattering configuration for the two-component BEC. 
(b) Bloch sphere representation of the magnetization of the BEC and the pseudomagnetic field $\vc{B}$ within the barrier.
(c) Spin rotation, $S_x$ (orange), $S_y$ (green), and $S_z$ (magenta), during collision of right-going (dashed curves) and left-going (solid curves) atomic density (black), generated in reflection from a barrier that also acts as a magnetic field. Initially, at $t=t_0$, the spin of the right-going wavepacket is polarized along $\ket{+x}$. Later, reflection from the barrier generates a left-going wavepacket whose spin, not parallel with that of the right-going wave, rotates about the right-going spin (back-action on the right-going spin is not illustrated for simplicity). After complete reflection, at $t=t_f$, a spin texture is evident across the left-going density.
}
\end{figure}

We study the generation of spin textures in a BEC of $^{87}\text{Rb}$ atoms in the $m_F=0$ ``clock'' states of the $F=1$ and $F=2$ ground-state hyperfine manifolds, where $F$ is the total angular momentum. 
This system can be conceptualized as a pseudospin-half system, where $\ket{2,0}\equiv\ket{+x}$ and $\ket{1,0}\equiv\ket{-x}$ [Fig.~\ref{fig:setup}(a,b)]. Thanks to a coincidence of scattering lengths in $^{87}\text{Rb}$, atom-atom interactions can be treated as spin-independent \cite{SpinorStamper}. 
The experimental setup has been described in detail previously \cite{Ramos2020,spierings2021fast}. 
Typically, $3\times 10^3$ atoms in a nearly pure condensate are prepared in $\ket{x}$ with a rms velocity width reduced to $\approx$\SI{0.3}{mm\per\s} via delta-kick cooling. Atoms are accelerated by a variable-duration magnetic-field gradient, and guided 
by a \SI{1054}{\nano\meter} optical beam to preserve quasi-1D motion along the $y$ axis. Typical clouds have a peak chemical potential $\mu/h=\SI{50}{Hz}$, kinetic energy $E_k/h = \SI{1.7}{kHz}$ (velocity \SI{3.9}{\mm\per\s}), and cloud length \SI{120}{\micro\m}. 

As illustrated in Fig.~\ref{fig:setup}(c), the atoms encounter, and are partially reflected from, a barrier. The barrier is a \SI{421.38}{\nano\meter} beam of light that has a Gaussian profile with $1/e^2$ radius of \SI{1.3}{\micro\m} along the $y$ direction. It is overlapped by a pair of resonant Raman beams 
which act as a pseudomagnetic field localized to the barrier region and pointing along the $z$ axis of the Bloch sphere [see Fig.~\ref{fig:setup}(b)].  Hence, reflection generates a counter-propagating matter wave whose spin is rotated to an extent controlled by the Raman Rabi frequency $\Omega_B$. 
These reflected atoms must then propagate through the rest of the condensate, and collisions occur between reflected and incident atoms. 
These collisions are observed to have a significant effect, whereas in the previously studied case of transmitted atoms \cite{Ramos2020,spierings2021fast}, they could be neglected. 
The phenomenon at the heart of our study is the counterflow spin dynamics, in which left-going spin rotates about right-going spin and vice versa. 
Note that the Raman beams are implemented using the same spatial mode as the barrier, through a combination of phase modulation at the clock frequency and attenuation of unwanted sidebands. 

\begin{figure*}[t] 
\includegraphics[width=\textwidth]{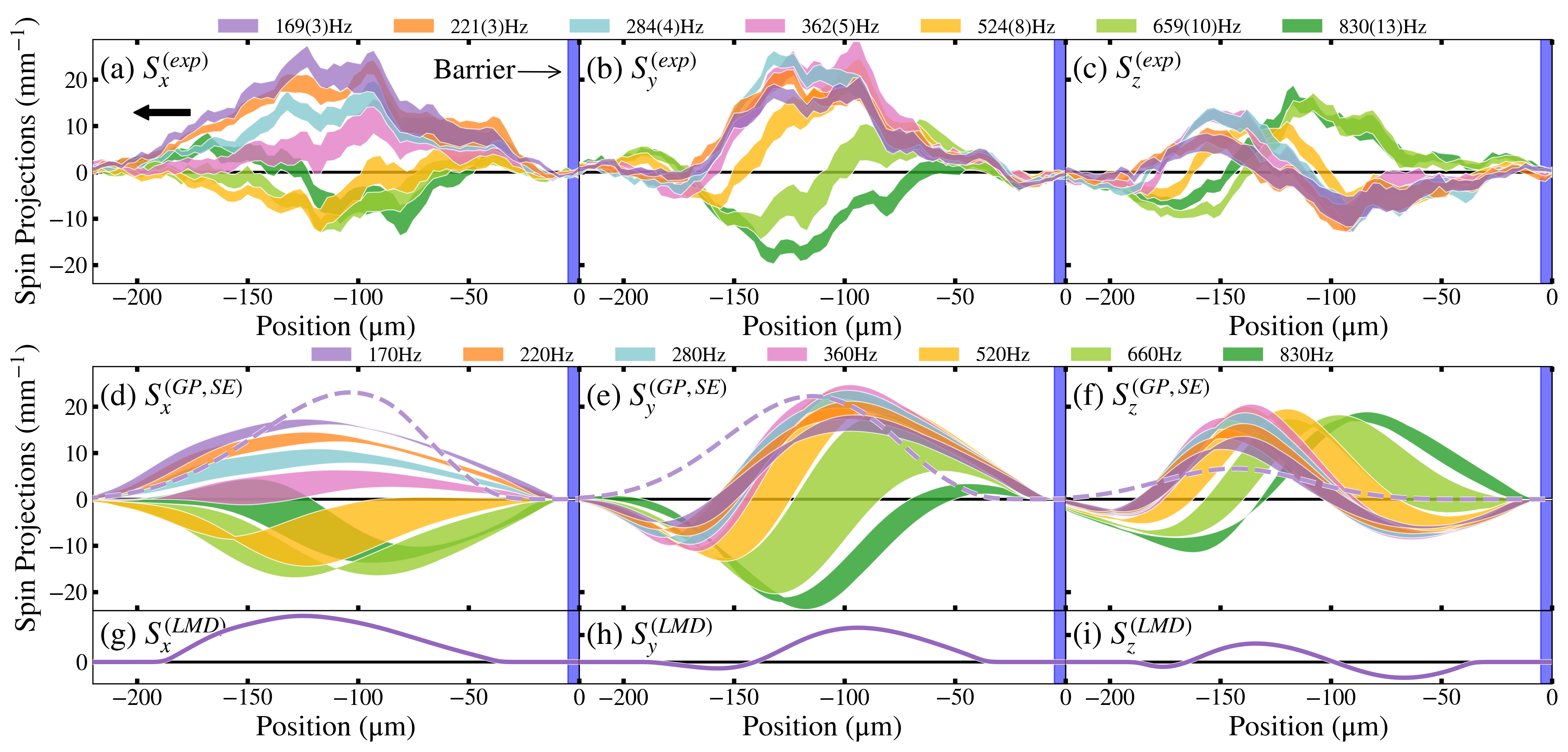}
\caption{{\bf Observed spin textures.} $S_x$ (left), $S_y$ (middle), $S_z$ (right) spatial profiles for the reflected atomic cloud for various $\Omega_B$ 
(denoted by color). The atomic cloud, with mean incident energy of \SI{1.67\pm0.05}{kHz} and rms energy width of \SI{0.30\pm0.03}{kHz}, scattered from a barrier, located at the origin, with peak energy of \SI{2.25\pm0.09}{kHz}. (a,b,c) Measured spin profiles with the bands representing one standard error in either direction from the average profile of the data set. (d,e,f) Spin profiles predicted by GP simulations of the experiment. The shaded regions are bounded by simulations including one standard error in the measured barrier height and velocity width. The dashed line shows the predicted spin profile without atomic interactions for $\Omega_B/2\pi = \SI{170}{Hz}$. (g,h,i) Predictions of the LMD model at $\Omega_B/2\pi=\SI{170}{Hz}$ for the average reflected velocity of this data set.\label{fig:profiles}}
\end{figure*}

After the wavepacket has left the barrier region, spin tomography is performed through a combination of sequential absorption imaging of the $\ket{\pm x}$ populations and a microwave pulse to rotate the axis of interest onto this  measurement basis \cite{spierings2021fast}. Spin profiles are extracted by computing the difference in the measured atom number in each absorption image pixel by pixel, integrating along the perpendicular direction, and dividing by the total reflected atom number. 
We report $S_i(y)=\int\! dx\, dz\, \sigma_i(\br)\rho(\br)/N_r$, where $\sigma_i$ is the $i$th component of the Bloch vector of the cloud, and $N_r$ is the total reflected atom number. We also report the aggregate rotation angles: $\theta_y$ describes the precession angle in the $S_x - S_y$ plane, and $\theta_z$ describes rotation out of that plane, resulting from the preferential reflection of the $\ket{+z}$ spin component which experiences an effectively higher barrier because it is parallel to the pseudomagnetic field \cite{buttiker1983larmor}.
 
Figure \ref{fig:profiles}(a,b,c) shows the spin profiles observed in the reflected cloud for a BEC incident with average energy well below the barrier height, $V_B$. For these data, $\sim 95\%$ of atoms are reflected. 
Because the magnetic barrier has a spin-dependent transmission, the reflected cloud is partially polarized along $+z$. In the absence of interactions, this polarization direction would be essentially constant across the cloud. 
A billiard-like picture of elastic collisions could allow the spin to be redistributed spatially, but would never lead to local polarization along $-z$.
Instead, we observe a full spin oscillation, which is a signature of coherent spin dynamics. We compare the observed spin texture to 1D GP, Figs.~\ref{fig:profiles}(d,e,f), and 1D LMD, Figs.~\ref{fig:profiles}(g,h,i), predictions. 

The two-component GP simulation uses equal interaction strengths between all spin states \cite{supp}, such that all atoms experience the same mean-field effective potential, $V_\text{eff}=g \rho$, proportional to the total density. However, the interaction energy can be different for the two spin components, since it depends on the overlap of the spatial wavefunction with the effective potential. This leads to mean-field-driven spin rotations generated by the difference in interaction energy between matter waves that do and do not interfere (i.e.~parallel vs.~anti-parallel spins). 

As shown in Fig.~\ref{fig:profiles}, the GP simulations capture many details of the observed spin textures. Without interactions, Schr\"{o}dinger equation (SE) profiles (dashed line in panels d,e,f) are smooth and positive for small $\Omega_B$. Aside from a small effect due to position-velocity correlations in our system, the SE predicts constant polarization across the cloud, even for larger $\Omega_B$. Interactions drive a spatially varying polarization, including sign changes in $S_z$ and $S_y$, whose positions are well captured by the GP simulations at various $\Omega_B$. 

The LMD model isolates the effective magnetic action of atomic collisions from the complicated density dynamics that occur during reflection of the cloud from the barrier. These calculations use only the average reflected velocity and a static density profile, and thus omit wavepacket spreading, density changes due to mean-field kinematics during reflection, as well as velocity-position correlations in predicting the spin texture. Reflected spins traverse the trailing portion of the atomic cloud after reaching the barrier position and experiencing a spin-dependent reflectivity determined by the SE simulation.

As shown in Fig.~\ref{fig:profiles}(g,h,i), the LMD captures the low-$\Omega_B$ ($\hbar \Omega_B/2 \ll V_B-E_k$) spin texture, which can be understood as follows. The barrier rotation is a small angle; the first atoms reflected rotate about the incoming $\ket{+x}$ spins. We see from the figure that this results in a sign flip in $S_y$ and two sign flips in $S_z$ (compared to the SE prediction), i.e., a relative phase shift of $\sim \pi$ between the $\ket{+x}$ and $\ket{-x}$ components. This is consistent with a transit time of roughly 10\,ms and $\mu/h \sim 50$\,Hz. Since the collisions conserve the net magnetization, the integrals of the profiles agree with the SE prediction in the small-$\Omega_B$ limit of Fig.~\ref{fig:profiles}; this explains the emergence of a spin texture with a sign flip.

\begin{figure}[t]
\includegraphics[width=\columnwidth]{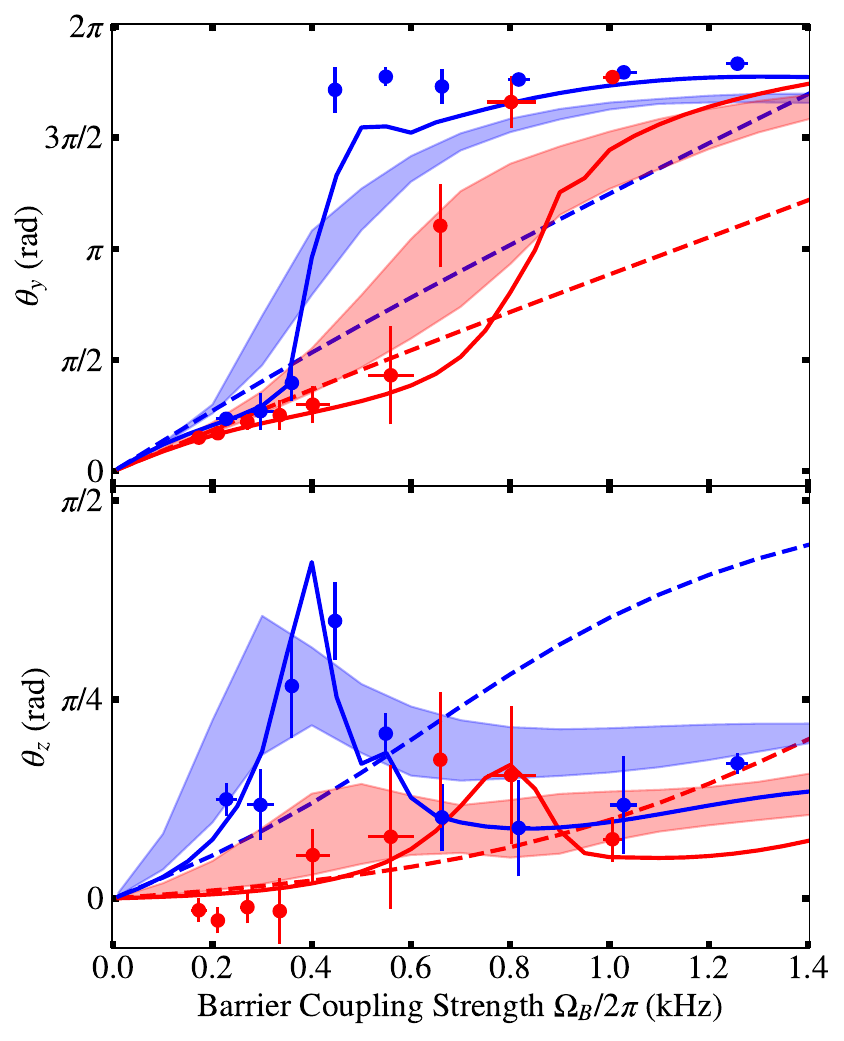}
\caption{{\bf Net rotation angles of the reflected cloud.} 
\label{fig:angles} Angles of rotation, $\theta_y$ (top) and $\theta_z$ (bottom), of the reflected atomic cloud for incident wavepackets well below the barrier height (red) and near the barrier height (blue). The peak energies of the barriers are \SI{2.25\pm0.09}{kHz} and \SI{1.72\pm0.07}{kHz} in the two cases, while the incident energy of the cloud was \SI{1.61\pm0.06}{kHz} with an rms width of \SI{0.29\pm0.03}{kHz}. Markers represent the measured angles of rotation. Color-coded bands show GP simulations of the experiment, with the shaded regions bounded as in Fig.~\ref{fig:profiles}, while the dashed lines indicate SE simulations for the average parameters. The solid curves depict LMD calculations for the average reflected velocity in each scenario.
}
\end{figure}

Let us now consider the regime of larger $\Omega_B$, where we can no longer make the assumption that the difference between the incident and reflected polarizations is small. 
Absent atomic interactions, $\theta_y$ will continue to precess about the pseudomagnetic field and when $\hbar \Omega_B$ becomes comparable to the energy deficit of the incident particles with respect to the barrier height, $\theta_z$ will tend toward $\pi/2$ as the reflected spin becomes polarized along the field direction.

Figure \ref{fig:angles} shows $\theta_y$ and $\theta_z$ measured for variable $\Omega_B$ and two barrier heights. 
We observe that both angles flatten out as a function of $\Omega_B$, in qualitative agreement with the GP and LMD calculations. Once the reflected spin becomes polarized along $+z$, trailing atoms are driven toward $\ket{+z}$ {\it prior} to reflection from the barrier \cite{supp}. Thus, above a certain $\Omega_B$, the net spin at the end of the scattering event is largely determined by atomic collisions rather than rotation about the external $\Omega_B$ drive. We attribute the discrepancy between the angles at which the dynamics saturate in the models and the experimental data to the precise details of the atomic density during the collision with the barrier that are not captured by the idealized 1D simulations. Note that since the LMD calculation does not incorporate the interactions between atoms at different velocities, it cannot replicate the spatial spin profiles in the large-$\Omega_B$ regime where the spin-dependent reflectivity becomes more sensitive to incident energy \cite{supp}. Nevertheless, the aggregate angles agree well even at higher $\Omega_B$ because they are predominantly set by the total spin rotation atoms acquire while traversing the atomic cloud. 

\begin{figure}[t]
\includegraphics[width=\columnwidth]{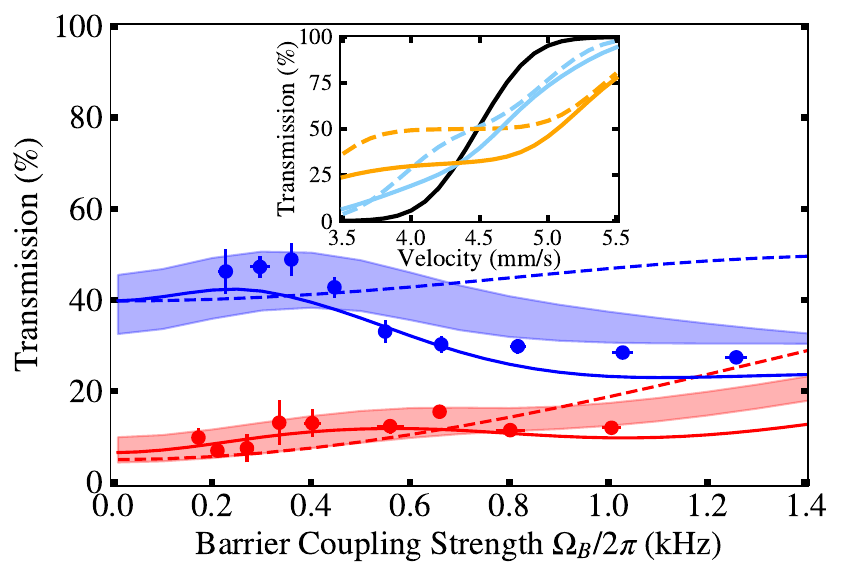}
\caption{{\bf Effect of spin rotation on transmission.} \label{fig:transmission}Transmission for the data presented in Fig.~\ref{fig:angles}, color-coded as before. Markers display data with statistical error bars, shaded bands illustrate GP simulations bounded as in previous plots, and the dashed lines depict SE simulations for the average experimental parameters. The solid lines indicate LMD calculations averaged over the \SI{0.29}{kHz} rms energy width of the incident cloud.  Inset) Transmission versus velocity for instances with $\Omega_B/2\pi$ of \SI{0}{\kilo\hertz} (black), \SI{1}{\kilo\hertz} (blue), and \SI{2}{\kilo\hertz} (orange). Dashed lines depict SE simulations, while solid lines represent GP simulations. In the inset, the simulated energy width is \SI{0.18}{kHz}, narrower than in the experiment to demonstrate the trends clearly.
}
\end{figure}

Transmission, shown in Fig.~\ref{fig:transmission}, corroborates that spin rotations occur prior to collision with the barrier. When the spin of the trailing portion of the cloud is driven toward $\ket{+z}$, reflection is enhanced because this state experiences an effectively higher barrier. The SE predicts that transmission would tend toward $50\%$ for large $\Omega_B$ because all atoms collide with the barrier in $\ket{+x}=(\ket{+z}+\ket{-z})/\sqrt{2}$. Without atomic interactions, as $|\theta_z|\rightarrow \pi/2$, transmission flattens at $50\%$ over a range of incident velocities, as exhibited in the inset of Fig.~\ref{fig:transmission}. Instead, for energies near the barrier height, we observe enhanced reflection at large $\Omega_B$ in qualitative agreement with the simulations incorporating atomic interactions.

Returning to Fig.~\ref{fig:angles}, we note that  $\theta_y$ and $\theta_z$ become $\Omega_B$-insensitive at lower $\Omega_B$ for the lower of the two barriers shown. 
For large energy deficits below the barrier, the time reflected atoms spend inside the barrier is expected to be shorter than for energies closer to the barrier height and hence for fixed $\Omega_B$ the rotation angles expected to be smaller in the former case (as indicated by the SE predictions in Fig.~\ref{fig:angles}). This is consistent with previous observations that transmitted particles spend less time interacting with the barrier for lower incident energies \cite{spierings2021fast}. Yet, we do not associate the observed angles with the time reflected atoms spend in the barrier because spin rotations occur prior to interaction with the barrier. While conservation of angular momentum during atomic collisions does preserve the net angles of the cloud, premature rotations cause the spin direction of many atoms entering the barrier to be unknown. 

In conclusion, we observe spin dynamics in a two-component BEC with spin-independent interactions. We show that in a counterflow scenario, four-wave mixing gives rise to magnetodynamics. 
In a mean-field picture, this system behaves as a phase-coherent 
two-component fluid, and spin rotations are caused by the interaction energy difference between components that do and do not experience interference. 
This is a new example of a spin-rotation effect for identical spins, closely related to ISRE. 
In both effects, the spin degree of freedom serves as a distinguishing particle label. In the case of ISRE it is the interaction energy due to particle exchange that is present only for identical spins, while here it is the density fluctuations due to single-particle interference -- which also occurs only for identical spins.
Their common physical origin is underscored by the fact that in both cases, changing either the particle statistics, bosons $\leftrightarrow$ fermions, or the sign of the interactions, repulsive $\leftrightarrow$ attractive, would reverse the direction of spin rotation \cite{supp}. 

\begin{acknowledgments}
We acknowledge Jeff McGuirk, Dan Stamper-Kurn, Joseph McGowan, Nick Mantella, and Harshil Neeraj for helpful discussions. This work was supported by NSERC, the Fetzer Franklin Fund of the John E. Fetzer Memorial Trust, and AFOSR FA9550-19-1-0365. DCS acknowledges support from the Mitacs Accelerate program. AMS is a Fellow of CIFAR. 
\end{acknowledgments}

\bibliography{bibliography} 

%%%%%%%%%% Merge with supplemental materials %%%%%%%%%%
\onecolumngrid
\clearpage
\begin{center}
\textbf{\large Supplemental Material: Spin Rotations in a Bose-Einstein Condensate Driven by Counterflow and Spin-independent Interactions}
\end{center}
%%%%%%%%%% Merge with supplemental materials %%%%%%%%%%
%%%%%%%%%% Prefix a "S" to all equations, figures, tables and reset the counter %%%%%%%%%%
\setcounter{equation}{0}
\setcounter{figure}{0}
\setcounter{table}{0}
\setcounter{page}{1}
\makeatletter
\renewcommand{\theequation}{S\arabic{equation}}
\renewcommand{\thefigure}{S\arabic{figure}}
\renewcommand{\bibnumfmt}[1]{[S#1]}
% \renewcommand{\citenumfont}[1]{S#1}
%%%%%%%%%% Prefix a "S" to all equations, figures, tables and reset the counter %%%%%%%%%%

\section{Local Magnetodyanmic Model for Two Counter-propagating Modes\label{sec:sup_LMD_derivation}}
The Hamiltonian for particles with two-body s-wave, contact interactions is given by $\hat{H}=\hat{H}_0+\hat{H}_\text{int}$, with
\begin{equation}
    \hat{H}_0=\int \! d{\bf x}\, \hat{\psi}^\dagger({\bf x}) \bigg(-\frac{\hbar^2}{2m}\nabla^2 + V({\bf x} )\bigg)\hat{\psi}({\bf x})\qquad\text{and}\qquad \hat{H}_\text{int}=\frac{g}{2} \int\! d{\bf x}\,\hat{\psi}^\dagger({\bf x})\hat{\psi}^\dagger({\bf x})\hat{\psi}({\bf x})\hat{\psi}({\bf x}).
\end{equation}
Here $g=4\pi\hbar^2a/m$ and $\hat{\psi}^\dagger({\bf x})=\sum_i \hat{a}^\dagger_i\phi^*_i({\bf x})$ is a field operator that creates a particle, via mode operator $\hat{a}^\dagger_i$, in the spatial mode $\phi^*_i({\bf x})$, where the set $\{\phi_i({\bf x})\}$ form a complete orthonormal basis. We consider only the evolution due to $\hat{H}_\text{int}$, so that the Heisenberg equation of motion for $\hat{\psi}({\bf x})$ is
\begin{equation}
    i\hbar\frac{\partial\hat{\psi}({\bf x})}{\partial t}=\big[\hat{\psi}({\bf x}),\hat{H}_\text{int}\big]=g\hat{\rho}({\bf x})\hat{\psi}({\bf x})\label{eq:eom},
\end{equation}
where $\hat{\rho}({\bf x})=\hat{\psi}^\dagger({\bf x})\hat{\psi}({\bf x})$ is the density operator. 

\subsection{Two-mode approximation}

Motivated by the experimental situation in which counter-propagating matter waves traverse one another, we consider a mode-space restricted to just two modes, labelled $+$ and $-$. Then $\hat{\psi}({\bf x})=\hat{a}_+\phi_+({\bf x}) + \hat{a}_-\phi_-({\bf x})$ and the density operator can then be written out as
\begin{align}
    \hat{\rho}({\bf x})&=\bigg(\hat{a}^\dagger_+\phi^*_+({\bf x})+\hat{a}^\dagger_-\phi^*_-({\bf x})\bigg)\bigg(\hat{a}_+\phi_+({\bf x})+\hat{a}_-\phi_-({\bf x})\bigg) \\
    &=\hat{\bar{n}}({\bf x})+\hat{a}^\dagger_+\hat{a}_-\phi^*_+({\bf x})\phi_-({\bf x})+\hat{a}^\dagger_-\hat{a}_+\phi^*_-({\bf x})\phi_+({\bf x})\label{eq:den},
\end{align}
where $\hat{\bar{n}}({\bf x})\equiv\hat{a}^\dagger_+\hat{a}_+\phi^*_+({\bf x})\phi_+({\bf x})+\hat{a}^\dagger_-\hat{a}_-\phi^*_-({\bf x})\phi_-({\bf x})$ measures the density due to occupation within each mode, while the last two terms in Eq.~\eqref{eq:den} indicate interference between the two modes. Evaluating Eq.~\eqref{eq:eom},
\begin{align}
    i\hbar\bigg(\phi_+({\bf x})\frac{\partial\hat{a}_+}{\partial t}+\phi_-({\bf x})\frac{\partial\hat{a}_-}{\partial t}\bigg)&=g\bigg(\hat{\bar{n}}({\bf x})+\hat{a}^\dagger_+\hat{a}_-\phi^*_+({\bf x})\phi_-({\bf x})+\hat{a}^\dagger_-\hat{a}_+\phi^*_-({\bf x})\phi_+({\bf x})\bigg)\bigg(\hat{a}_+\phi_+({\bf x}) + \hat{a}_-\phi_-({\bf x})\bigg) \\
    &=\begin{aligned}[t]&g\hat{\bar{n}}({\bf x})\bigg(\hat{a}_+\phi_+({\bf x}) + \hat{a}_-\phi_-({\bf x})\bigg)\\ &+g\begin{aligned}[t]&\bigg(\hat{a}^\dagger_+\hat{a}_-\hat{a}_+|\phi_+({\bf x})|^2\phi_-({\bf x})+\hat{a}^\dagger_-\hat{a}_+\hat{a}_+\phi_+({\bf x})\phi^*_-({\bf x})\phi_+({\bf x}) \\
    & + \hat{a}^\dagger_+\hat{a}_-\hat{a}_-\phi_-({\bf x})\phi^*_+({\bf x})\phi_-({\bf x})+\hat{a}^\dagger_-\hat{a}_+\hat{a}_-|\phi_-({\bf x})|^2\phi_+({\bf x})\bigg).\end{aligned}\end{aligned}\label{eq:ortho}
\end{align}
This equation of motion can be separated into equations describing the evolution of each mode individually by course-graining over position. For example, consider plane-wave spatial modes, $\phi_\pm({\bf x})=V^{-1/2}\exp{\pm i{\bf k\cdot x}}$, which have momentum $\pm\hbar {\bf k}$ and $|\phi_\pm({\bf x})|^2=1$ (suppressing normalization constants, which we do for simplicity below). Multiplying Eq.~\eqref{eq:ortho} by $\phi^*_+({\bf x})$ or $\phi^*_-({\bf x})$ and integrating over all space, the orthogonality of $\phi_\pm({\bf x})$ provides the simple evolution equations for $\hat{a}_+$ and $\hat{a}_-$, respectively
\begin{equation}
    i\hbar\frac{\partial\hat{a}_+}{\partial t}=g\big(\hat{a}^\dagger_+\hat{a}_++\hat{a}^\dagger_-\hat{a}_-)\hat{a}_++g\hat{a}^\dagger_-\hat{a}_+\hat{a}_-\qquad\qquad i\hbar\frac{\partial\hat{a}_-}{\partial t}=g\big(\hat{a}^\dagger_+\hat{a}_++\hat{a}^\dagger_-\hat{a}_-)\hat{a}_-+g\hat{a}^\dagger_+\hat{a}_-\hat{a}_+.\label{eq:4wave_mixing}
\end{equation}
The terms in parentheses come from $\hat{\bar{n}}({\bf x})$ and correspond to the evolution of each mode due to the average occupation of $\phi_\pm({\bf x})$. The final terms, on the other hand, indicate wave mixing between the $+$ and $-$ modes without changing the population in each mode. All position dependence has been removed from Eq.~\eqref{eq:4wave_mixing}, because the course-graining was done over all space. But, the same simplification can be made for creation operators of `momentum side modes,' $\hat{a}^\dagger_m({\bf k})=\int d{\bf x}\varphi_m^*({\bf x})\exp({i{\bf k\cdot x}})\hat{\psi}^\dagger({\bf x})$ of general spatial wavefunctions, $\varphi_m({\bf x})$, that have finite extent. Two momentum side modes with spatial dependence $\varphi_+({\bf x})\exp({i{\bf k_+\cdot x}})$ and $\varphi_-({\bf x})\exp({i{\bf k_-\cdot x}})$ for ${\bf k_+}\neq{\bf k_-}$ are approximately orthogonal for regions large compared to $\Delta{\bf x}\equiv|2\pi/({\bf k_+} - {\bf k_-})|$ so long as $\varphi_\pm({\bf x})$ vary negligibly over $\Delta {\bf x}$ in each dimension. In this case, position is course-grained over $\Delta {\bf x}$ and the densities given by $\varphi_\pm({\bf x})$ are approximated to be locally constant within $\Delta {\bf x}$. In more physical terms, this approximation imposes a `local' conservation of momentum, in addition to the conservation of momentum for the entire, isolated system. When $\varphi_\pm({\bf x})$ vary negligibly over $\Delta {\bf x}$, the momentum set by ${\bf k_\pm}$ is much greater than any other momenta incorporated in $\varphi_\pm({\bf x})$. As a result, elastic collisions cannot change the occupation of each momentum side mode even locally and such terms drop out of the evolution when integrated over regions larger than $\Delta {\bf x}$, as indicated in Eq.~\eqref{eq:4wave_mixing}. What remains is wave mixing due to collisions that maintains the occupation of each mode.

\subsection{Interaction-driven spin dynamics}

In order to see the impact of this approximation on spin rotations, we must include a spin degree of freedom. We indicate the pseudospin-half of the experiment via $\uparrow$ and $\downarrow$. With nearly equal interaction strengths due to a coincidence of scattering lengths in $^{87}\text{Rb}$, $g_{\uparrow\uparrow}\approx g_{\downarrow\downarrow}\approx g_{\uparrow\downarrow}\equiv g$ and spin interchange in collisions is negligible. We assume that the field operators for each spin state can be written in terms of the same plane-wave spatial modes, $\phi_\pm({\bf x})$, so that the mode operators for each spin state evolve according to Eqs.~\eqref{eq:4wave_mixing}, with the density summed over both spin states. This implies that no evolution occurred prior to the evolution under $\hat{H}_\text{int}$ that populated modes other than $\pm$ for either spin state. Additionally, below we consider evolution along one spatial dimension. Considering only the $-$ mode, we have
\begin{equation}
    i\hbar\frac{\partial}{\partial t} \begin{bmatrix} \hat{a}_{-,\uparrow} \\ \hat{a}_{-,\downarrow} \end{bmatrix} = g\big(\hat{\bar{n}}_\uparrow+\hat{\bar{n}}_\downarrow\big)\begin{bmatrix} \hat{a}_{-,\uparrow} \\ \hat{a}_{-,\downarrow} \end{bmatrix}+g\big(\hat{a}^\dagger_{+,\uparrow}\hat{a}_{-,\uparrow}+\hat{a}^\dagger_{+,\downarrow}\hat{a}_{-,\downarrow}\big)\begin{bmatrix} \hat{a}_{+,\uparrow} \\ \hat{a}_{+,\downarrow} \end{bmatrix},\label{eq:4wave_mixing_spin}
\end{equation}
where $\hat{\bar{n}}_\sigma\equiv\hat{n}_{+,\sigma}+\hat{n}_{-,\sigma}\equiv\hat{a}^\dagger_{+,\sigma}\hat{a}_{+,\sigma}+\hat{a}^\dagger_{-,\sigma}\hat{a}_{-,\sigma}$ for $\sigma=\uparrow,\downarrow$. As before, the first term corresponds to evolution due to the average density of each mode, and the later term represents wave mixing, transferring population between states with momentum $\pm\hbar k$. Here, we see that for the pseudospin states with momentum $-\hbar k$ Bragg scattering from the density modulation, produced by mixing of the $+$ and $-$ modes within each spin state, scatters a particle from $-$ to $+$ without flipping its spin.

Let us expand Eq.~\eqref{eq:4wave_mixing_spin} to elucidate the spin rotation effect. First, note that the far right term can be rearranged as
\begin{align}
\big(\hat{a}^\dagger_{+,\uparrow}\hat{a}_{-,\uparrow}+\hat{a}^\dagger_{+,\downarrow}\hat{a}_{-,\downarrow}\big)\begin{bmatrix} \hat{a}_{+,\uparrow} \\ \hat{a}_{+,\downarrow} \end{bmatrix} = \begin{bmatrix} \hat{a}^\dagger_{+,\uparrow}\hat{a}_{-,\uparrow}\hat{a}_{+,\uparrow} + \hat{a}^\dagger_{+,\downarrow}\hat{a}_{-,\downarrow}\hat{a}_{+,\uparrow}\\ \hat{a}^\dagger_{+,\uparrow}\hat{a}_{-,\uparrow}\hat{a}_{+,\downarrow} + \hat{a}^\dagger_{+,\downarrow}\hat{a}_{-,\downarrow}\hat{a}_{+,\downarrow} \end{bmatrix}&=\pm\begin{bmatrix} \hat{a}^\dagger_{+,\uparrow}\hat{a}_{+,\uparrow}\hat{a}_{-,\uparrow} + \hat{a}^\dagger_{+,\downarrow}\hat{a}_{+,\uparrow}\hat{a}_{-,\downarrow}\\ \hat{a}^\dagger_{+,\uparrow}\hat{a}_{+,\downarrow}\hat{a}_{-,\uparrow} + \hat{a}^\dagger_{+,\downarrow}\hat{a}_{+,\downarrow}\hat{a}_{-,\downarrow} \end{bmatrix}\\
&=\pm\begin{bmatrix} \hat{n}_{+,\uparrow} & \hat{a}^\dagger_{+,\downarrow}\hat{a}_{+,\uparrow} \\ \hat{a}^\dagger_{+,\uparrow}\hat{a}_{+,\downarrow} & \hat{n}_{+,\downarrow} \end{bmatrix} \begin{bmatrix} \hat{a}_{-,\uparrow} \\ \hat{a}_{-,\downarrow} \end{bmatrix}
\end{align}
% \begin{align}
% \big(\hat{a}^\dagger_{+,\uparrow}\hat{a}_{-,\uparrow}+\hat{a}^\dagger_{+,\downarrow}\hat{a}_{-,\downarrow}\big)\begin{bmatrix} \hat{a}_{+,\uparrow} \\ \hat{a}_{+,\downarrow} \end{bmatrix} = \begin{bmatrix} \hat{a}^\dagger_{+,\uparrow}\hat{a}_{-,\uparrow}\hat{a}_{+,\uparrow} + \hat{a}^\dagger_{+,\downarrow}\hat{a}_{-,\downarrow}\hat{a}_{+,\uparrow}\\ \hat{a}^\dagger_{+,\uparrow}\hat{a}_{-,\uparrow}\hat{a}_{+,\downarrow} + \hat{a}^\dagger_{+,\downarrow}\hat{a}_{-,\downarrow}\hat{a}_{+,\downarrow} \end{bmatrix}&=\begin{bmatrix} \hat{a}^\dagger_{+,\uparrow}\overset{\longleftrightarrow}{\hat{a}_{+,\uparrow}\hat{a}_{-,\uparrow}} + \hat{a}^\dagger_{+,\downarrow}\overset{\longleftrightarrow}{\hat{a}_{+,\uparrow}\hat{a}_{-,\downarrow}}\\ \hat{a}^\dagger_{+,\uparrow}\underset{\longleftrightarrow}{\hat{a}_{+,\downarrow}\hat{a}_{-,\uparrow}} + \hat{a}^\dagger_{+,\downarrow}\underset{\longleftrightarrow}{\hat{a}_{+,\downarrow}\hat{a}_{-,\downarrow}} \end{bmatrix}\\
% &=\begin{bmatrix} \hat{n}_{+,\uparrow} & \hat{a}^\dagger_{+,\downarrow}\hat{a}_{+,\uparrow} \\ \hat{a}^\dagger_{+,\uparrow}\hat{a}_{+,\downarrow} & \hat{n}_{+,\downarrow} \end{bmatrix} \begin{bmatrix} \hat{a}_{-,\uparrow} \\ \hat{a}_{-,\downarrow} \end{bmatrix}
% \end{align}
In this rearrangement, $\hat{a}_{-,\sigma}$ appears to evolve according to density in the state with $+\hbar k$ momentum. Note that the $\pm$ in front is determined by the commutation(anti-commutation) relation for bosonic(fermionic) mode operators. Inserting this result into Eq.~\eqref{eq:4wave_mixing_spin} yields
\begin{align}
    i\hbar\frac{\partial}{\partial t} \begin{bmatrix} \hat{a}_{-,\uparrow} \\ \hat{a}_{-,\downarrow} \end{bmatrix} &= g\big(\hat{\bar{n}}_\uparrow+\hat{\bar{n}}_\downarrow\big)\begin{bmatrix} \hat{a}_{-,\uparrow} \\ \hat{a}_{-,\downarrow} \end{bmatrix}\pm g\begin{bmatrix} \hat{n}_{+,\uparrow} & \hat{a}^\dagger_{+,\downarrow}\hat{a}_{+,\uparrow} \\ \hat{a}^\dagger_{+,\uparrow}\hat{a}_{+,\downarrow} & \hat{n}_{+,\downarrow} \end{bmatrix} \begin{bmatrix} \hat{a}_{-,\uparrow} \\ \hat{a}_{-,\downarrow} \end{bmatrix} \\
    &=g\Bigg( \big(\hat{\bar{n}}_\uparrow+\hat{\bar{n}}_\downarrow\big)\mathds{1} \pm \begin{bmatrix} \hat{n}_{+,\uparrow} & 0 \\ 0 & \hat{n}_{+,\downarrow} \end{bmatrix} \pm \begin{bmatrix} 0 & (\hat{a}^\dagger_{+,\uparrow}\hat{a}_{+,\downarrow})^\dagger \\ \hat{a}^\dagger_{+,\uparrow}\hat{a}_{+,\downarrow} & 0 \end{bmatrix} \Bigg)\begin{bmatrix} \hat{a}_{-,\uparrow} \\ \hat{a}_{-,\downarrow} \end{bmatrix} \\
    &=g\Bigg( \big(\hat{\bar{n}}_\uparrow+\hat{\bar{n}}_\downarrow \pm \frac{1}{2}\hat{n}_{+,\uparrow} \pm \frac{1}{2}\hat{n}_{+,\downarrow})\mathds{1} \pm \frac{1}{2}(\hat{n}_{+,\uparrow} - \begin{aligned}[t]&\hat{n}_{+,\downarrow}){\bf \sigma_3} \pm \frac{1}{2}(\hat{a}^\dagger_{+,\uparrow}\hat{a}_{+,\downarrow}+(\hat{a}^\dagger_{+,\uparrow}\hat{a}_{+,\downarrow})^\dagger){\bf \sigma_1} \\ &\pm \frac{-i}{2}(\hat{a}^\dagger_{+,\uparrow}\hat{a}_{+,\downarrow}-(\hat{a}^\dagger_{+,\uparrow}\hat{a}_{+,\downarrow})^\dagger){\bf \sigma_2} \Bigg)\begin{bmatrix} \hat{a}_{-,\uparrow} \\ \hat{a}_{-,\downarrow} \end{bmatrix}\end{aligned} \\
    &=g\Bigg( \big(\hat{\bar{n}}_\uparrow+\hat{\bar{n}}_\downarrow \pm \frac{1}{2}\hat{n}_{+,\uparrow} \pm \frac{1}{2}\hat{n}_{+,\downarrow}\big)\mathds{1} \pm \frac{1}{2}\big(\hat{s}_x{\bf \sigma_1} + \hat{s}_y{\bf \sigma_2} + \hat{s}_z{\bf \sigma_3}\big) \Bigg)\begin{bmatrix} \hat{a}_{-,\uparrow} \\ \hat{a}_{-,\downarrow} \end{bmatrix} \\
    &\equiv\hat{H}_\text{mag}\begin{bmatrix} \hat{a}_{-,\uparrow} \\ \hat{a}_{-,\downarrow} \end{bmatrix},
\end{align}
where ${\bf \sigma_1},{\bf \sigma_2},{\bf \sigma_3}$ are Pauli matrices, and $\hat{s}_i$ is the operator which measures the $i$th component of the Bloch vector, for $i=x,y,z$. Thus, taking the expectation value of the effective Hamiltonian $\hat{H}_\text{mag}$ for any pseudospin-half system $\ket{\Psi}$, we see explicitly a magnetic action
\begin{equation}
    \bra{\Psi}\hat{H}_\text{mag}\ket{\Psi}=g\big(\frac{2\pm1}{2}\langle\hat{n}_+\rangle+\langle\hat{n}_-\rangle\big)\mathds{1} \pm \frac{1}{2}g\langle\hat{n}_+\rangle\langle\hat{\vec{\sigma}}_+\cdot\vec{\sigma}\rangle \label{eq:expected_Hmag},
\end{equation}
where $\hat{n}_\pm=\sum_\sigma\hat{a}_{\pm,\sigma}^\dagger\hat{a}_{\pm,\sigma}$ measures the occupation of each momenta summed over spin state, $\hat{\vec{\sigma}}_+$ is a vector of $\hat{s}_i$ operators corresponding to the Bloch vector of the pseudospin-half in the $+$ mode, and $\vec{\sigma}$ is the Pauli vector. Equation \eqref{eq:expected_Hmag} has the form $A\mathds{1}+(\hbar/2)\vec{\Omega}\cdot\vec{\sigma}$, which leads to precession of $\vec{\sigma}$ about $\vec{\Omega}$ with frequency $\hbar|\vec{\Omega}|$ (i.e.~$\partial\vec{\sigma}/\partial t=\vec{\Omega}\times\vec\sigma$). Thus, the $-$ mode pseudospin experiences an effective magnetic interaction about the pseudospin of the $+$ mode with frequency $g\langle\hat{n}_+\rangle$. Following identical arguments for $\hat{a}_{+,\uparrow}$ and $\hat{a}_{+,\downarrow}$ leads to the corresponding conclusion that the $+$ mode pseudospin rotates about the $-$ mode pseudospin with frequency given by the density in $-$. In sum, the effective magnetic interaction is then given by
\begin{equation}
    \frac{\partial\vec{\sigma}_-}{\partial t}=\pm g\langle\hat{n}_+\rangle\vec{\sigma}_+\times\vec{\sigma}_- \qquad\text{and}\qquad \frac{\partial\vec{\sigma}_+}{\partial t}=\pm g\langle\hat{n}_-\rangle\vec{\sigma}_-\times\vec{\sigma}_+, \label{eq:effective_magnet_field}
\end{equation}
for the normalized Bloch vectors $\vec{\sigma}_\pm$ in each mode. Note that these arguments did not depend on whether $\ket{\Psi}$ represented an incoherent or coherent state, and so describe both nondegenerate and fully condensed systems. It is straightforward to extend these arguments to modes with finite extent, in which case Eq.~\eqref{eq:effective_magnet_field} has position dependence, as in the main text, which represents a course-graining of the density in each mode as described below Eq.~\eqref{eq:4wave_mixing}. In this scenario, we refer to this model as `local magnetodynamic' because it is only within the approximation that the average local density is constant with respect to the length scale set by counter-propagating modes that the magnetic interaction is an appropriate description. Lastly, the restriction to only two modes can be relaxed so long as the length scale over which the densities are coarse-grained satisfies the local density approximation for every pair of modes for both spin states. In this more general scenario, the Bloch vector of each mode will precess about the sum of many effective magnetic fields, $\vec{\Omega}_i=\pm g\langle n_i\rangle\vec{\sigma}_i$, for each mode $i$. In the modelling done for this experiment, however, we only ever consider two counter-propagating modes in one dimension.

\subsection{Validity of the Local Magnetodynamic Model in this experiment}

We prepare an atomic wavepacket from a quasi-1D BEC with a narrow momentum distribution that allows the interaction-driven spin dynamics to be well described by the local magnetodynamic model. The preparation uses delta-kick cooling (also known as matter-wave lensing), a well known shortcut to adiabatic expansion to reduce the momentum spread of the wavepacket while maintaining phase-space density. In an ideal implementation, an interacting BEC initially trapped in a one-dimensional harmonic trap, with trap frequency $\omega_0$, is released from the confining potential, allowed to expand for a time $t$, before a final harmonic potential is pulsed on to minimize the spread in velocity. Due to the scale-invariant dynamics of a BEC (in any spatial dimension in the Thomas-Fermi regime), the final cloud expands by a factor $b(t)=\sqrt{1+\omega_0^2t^2}$ while the momentum distribution is narrowed by the same factor. In this experiment, the cloud has a one-dimensional Thomas-Fermi radius, $R_\text{TF}$, after matter-wave lensing of about $\SI{60}{\micro\m}$. On the other hand, the center-of-mass (COM) velocity of the atomic wavepacket incident on the barrier is around $\SI{4}{\mm\per\s}$, which corresponds to a length scale for the difference in momenta of counter-propagating waves during reflection of about $\Delta x\approx\SI{0.6}{\micro\m}$. Thus, for the inverted parabolic shape of a one-dimensional Thomas-Fermi profile, $n(x)=1-x^2/R_\text{TF}^2$ for $|x|<R_\text{TF}$, the approximation of constant density over the length scale $\Delta x$ is well satisfied even at the edge of the cloud, where $|\frac{dn}{dx}\Delta x|=2\Delta x/R_\text{TF}\sim0.02$.

Due to experimental imperfections in the implementation of delta-kick cooling, the momentum width of the atomic cloud is not fully reduced by $b(t)$; the rms velocity width of the cloud is measured to be $\SI{0.35\pm0.03}{\mm\per\s}$. This width is ten times smaller than the average velocity of the accelerated cloud, such that the local conservation of momentum assumed in the course-grained approach is a good approximation. Put another way, this velocity variation corresponds to a full-cloud momentum width (i.e.~inverse thermal de Broglie wavelength) on the order of $h(\SI{5}{\micro\m})^{-1}$, compared to the standing-wave period of $\SI{0.6}{\micro\m}$. 

\section{Models for the experiment}
We compare our data to three models of the experiment governed by the 1D time-dependent Schr\"{o}dinger equation (SE), the 1D time-dependent Gross-Pitaevskii equation (GP), and a 1D local magnetodynamic model (LMD) derived in Sec.~\ref{sec:sup_LMD_derivation}. While the SE calculations model what the experiment might look like in the absence of atomic interactions, the GP simulations incorporate the influence of atomic interactions on both kinematics and spin dynamics, and the LMD calculations isolate the magnetic interaction of counter-propagating spins.

We perform GP simulations of the experiment using the following set of coupled equations: 
\begin{align}
\label{eq:gp1}
i\hbar\frac{\partial\psi_1}{\partial t}=\bigg[\frac{-\hbar^2}{2m}\nabla^2+V+g_{11}|\psi_1|^2+g_{12}|\psi_2|^2+\hbar\delta\bigg]\psi_1+\hbar\Omega\psi_2 \\
\label{eq:gp2}
i\hbar\frac{\partial\psi_2}{\partial t}=\bigg[\frac{-\hbar^2}{2m}\nabla^2+V+g_{22}|\psi_2|^2+g_{21}|\psi_1|^2\bigg]\psi_2+\hbar\Omega\psi_1,
\end{align}
where $V$ is the trapping and barrier potential, $\hbar\delta$ is an energy offset, $\Omega$ is the Rabi frequency of the Raman beams, and $g_{ij}=4\pi\hbar^2a_{ij}/m$ is the spin-dependent interaction parameter, with s-wave scattering length $a_{ij}$ between the $i$ and $j$ components. Given the aspect ratio of the guiding optical dipole trap (ODT), with $\omega_y<<\omega_x,\omega_z$, the dynamics of the atomic cloud are well described by Eqs.~\eqref{eq:gp1} and \eqref{eq:gp2} in reduced, one-dimensional form \cite{Bao2003}. Additionally, we set all scattering lengths equal to $96a_0$, for Bohr radius $a_0$, so that we do not observe spin dynamics induced by differences in interaction strengths. We use the time-splitting spectral method to solve the coupled one-dimensional GP equations. First, we use imaginary time propagation to find the ground state of the initial trapping configuration with an energy offset to initialize the wavepacket in spin state 1, which we associate with $\ket{x}$. We then release the atomic cloud from an initial crossed-trap and simulate matter-wave lensing similar to that done in the experiment. We adjust the duration of the lensing stage so that the simulated velocity width agrees with the measured value. This simulated wavepacket preparation is the initial condition for all of our models. 

For the GP simulations, the evolution of the wavepacket is modelled in the 1D longitudinal potential of the ODT and scatters from a Gaussian barrier with the measured spatial width. The incident velocity of the wavepacket is set by a phase gradient written across the initial wavepacket. We emulate the spin coupling inside the barrier by giving $\Omega(y)$ the same Gaussian spatial profile as the barrier and defining the effective Rabi frequency as in Sec.~\ref{sec:rabi_frequency}. These simulations not only capture spin dynamics due to the different interaction energies experienced by the two spin states, but also model the interplay between the repulsive interactions of the atoms and the confining potential of the ODT, which modify the velocity width and spatial shape of the wavepacket during collision with the barrier. The GP simulations are similar to those done in \cite{spierings2021fast}.

The SE calculations are performed like the GP simulations, except with the scattering length set to zero. Additionally, the longitudinal potential of the dipole trap is removed as otherwise this confining force would significantly alter the velocity distribution of the wavepacket in the absence of the competing repulsive atomic interactions.
 
\begin{figure*}[t] 
\includegraphics[width=\textwidth]{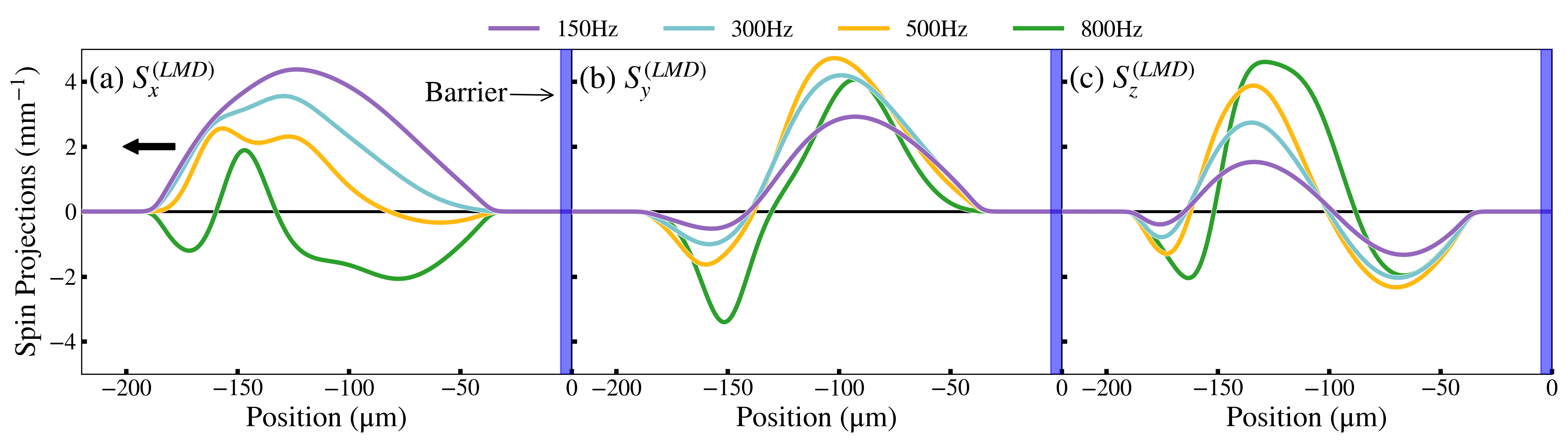}
\caption{{\bf Spin textures predicted by the LMD model.} $S_x$, $S_y$, and $S_z$ spatial profiles for the reflected atomic cloud for various $\Omega_B$ (denoted by color) for the same experimental parameters (mean incident energy of \SI{1.67\pm0.05}{kHz} and barrier height of \SI{2.25\pm0.09}{kHz}) as in Fig.~2 of the main text. The LMD calculations model the behavior of the average reflected velocity and the predicted spatial profiles begin to deviate from the experimental data and GP simulations when the rotation angles acquired in the barrier vary significantly across the rms energy width, \SI{0.30\pm0.03}{kHz}, of the wavepacket. Here, this is most notable when $E_k\sim V_B-\hbar\Omega_B/2$.\label{fig:sup_profiles}}
\end{figure*}

The LMD calculations assume a static density profile, given by the result of the simulated wavepacket preparation. Propagation is modelled by a monochromatic cloud incident on the barrier with velocity given by the average reflected velocity for the wavepacket in the SE calculations. Reflection from the barrier is imagined to instantaneously create counter-propagating density having had its spin manipulated (i.e.~an angle of rotation $\theta_y$ and angle of alignment $\theta_z$), and density reduced, by the barrier as anticipated by the SE given the incident spin state. Overlapping, counter-propagating density then experiences the magnetic interaction indicated by Eq.~\eqref{eq:effective_magnet_field} (see the provided animation for an illustrative example). In this way, the LMD calculations emulate the spin dynamics driven by atomic collisions in the experiment without the wavepacket expansion and other density dynamics that occur while atoms scatter from the Gaussian barrier. Accordingly, the LMD calculations do not capture the spatial spin dynamics in our experiment when the rotation angles acquired upon reflection from the pseudomagnetic barrier vary significantly across the velocity width of the atomic cloud, as demonstrated in Fig.~\ref{fig:sup_profiles}. The spin-dependent reflectivity of the barrier is most sensitive to the incident energy close to the barrier height. Hence, we see in Fig.~\ref{fig:sup_profiles} that the spin profiles begin to deviate from the experimental observations when the average incident kinetic energy, $E_k$, is comparable to the lower of the two dressed barrier heights, $V_B-\hbar\Omega_B/2$.

Nevertheless, the LMD model does show qualitative agreement for the net rotation angles in the reflected cloud regardless of $\Omega_B$, as seen in Fig.~3 of the main text. This is because at large $\Omega_B$ the final rotation angles are largely determined by the shape and density of the gas, as trailing spins become significantly rotated prior to reflection from the barrier. While small differences in incident energy can change the spin-dependent reflectivity of the barrier, they do not significantly change the traversal time of atoms counter-propagating through the cloud, which sets the interaction-driven rotation angle. In this experiment, the interaction-driven rotation angles are typically larger than those generated by the pseudomagnetic barrier at large $\Omega_B$. Thus, while the precise structure of the spin texture predicted by the LMD model at large $\Omega_B$ may differ from the experimental observation, the net rotation angles are largely the same.

\begin{figure*}[t]
    \centering
    \begin{minipage}[b]{0.49\textwidth}
        \includegraphics[width=\textwidth]{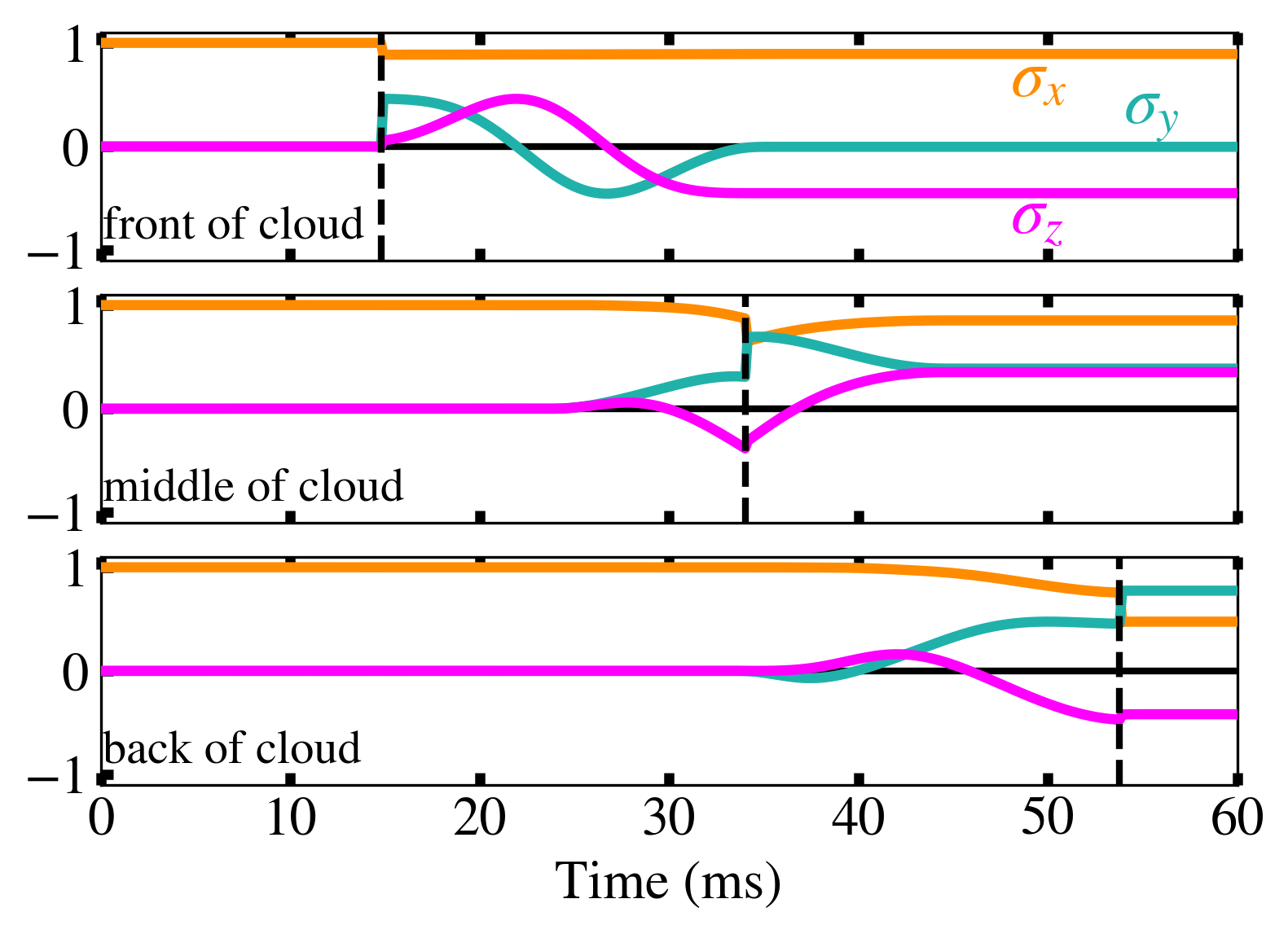}
        % \caption{Caption for Figure 1}
    \end{minipage}
    \hfill
    \begin{minipage}[b]{0.49\textwidth}
        \includegraphics[width=\textwidth]{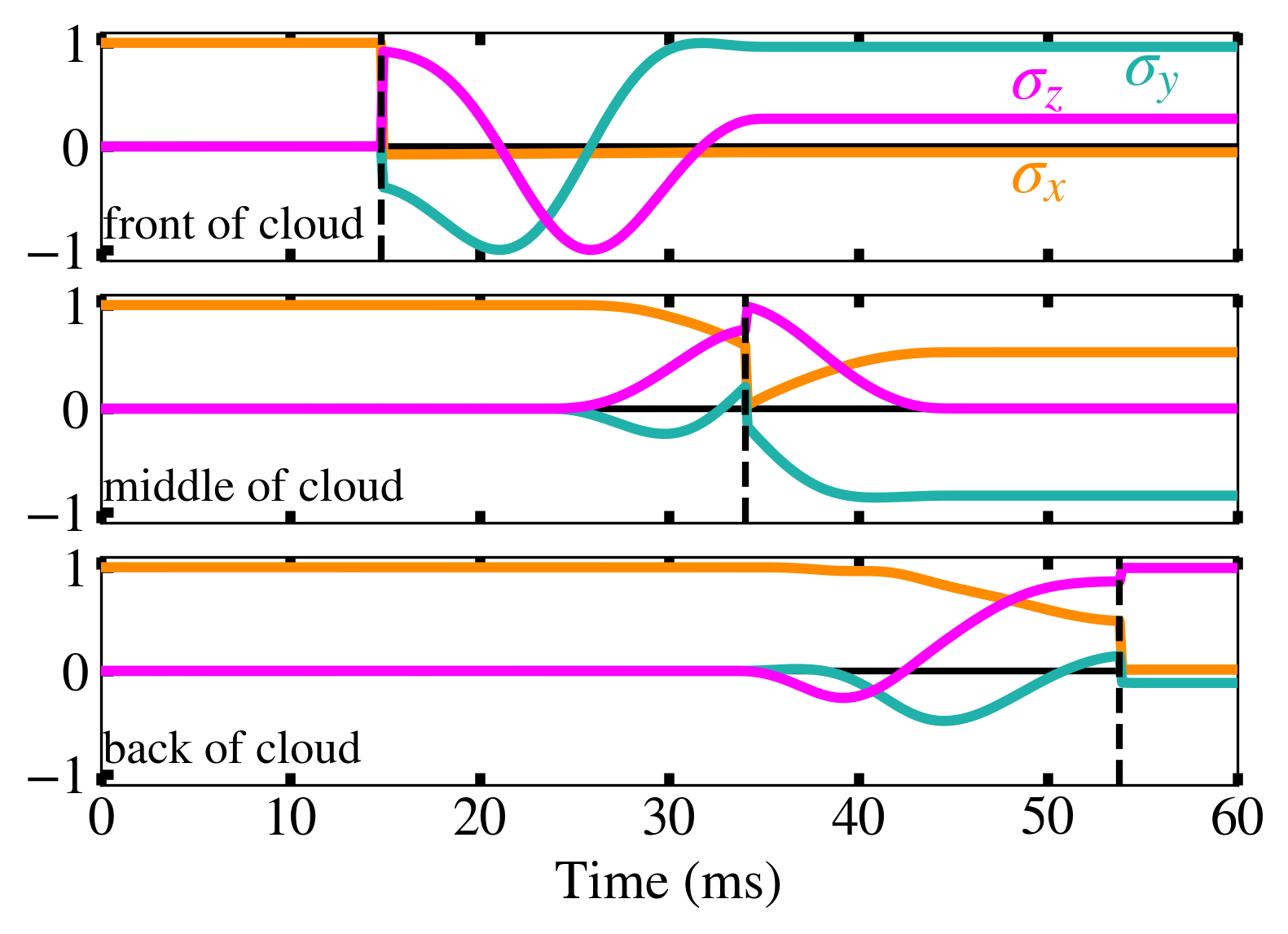}
        % \caption{Caption for Figure 2}
    \end{minipage}
    \caption{{\bf Magnetic rotations in the LMD model at given positions in the reflected cloud.} In the LMD model, atoms reflect from the barrier instantaneously, having acquired some spin rotation about the pseudomagnetic field, before traversing the trailing portions of the cloud. Here we show the $\sigma_x$, $\sigma_y$, $\sigma_z$ components of the Bloch vector at positions in the front, middle, and back edge of the cloud as they undergoing reflection from the barrier and the interaction-driven rotations. The barrier coupling strengths are (left) \SI{100}{\Hz} and (right) \SI{1000}{\Hz}. The vertical black dashed lines indicate the time at which a portion of the cloud reaches the barrier. The barrier height is \SI{1.72}{kHz} and the incident energy is \SI{1.51}{kHz}, matching the average reflected energy for the wavepacket incident on this barrier height in the experiment. Note that for large barrier coupling strength the middle and back of the cloud align toward $\ket{z}$ prior to collision with the barrier.}
    \label{fig::LMD_tracking}
\end{figure*}

The LMD model gives insight to spin rotations prior to interaction with the barrier since, given the fixed density profile in these calculations, one can track the spin components in different sections of the atomic cloud. For example, Fig.~\ref{fig::LMD_tracking} shows the evolution of the components of the Bloch vector in the front, middle, and back of the atomic cloud while the cloud reflects from the barrier. The spin components change instantaneously when each portion of the cloud interacts with the barrier. In addition, the effective magnetic rotations are evident as oscillations of the spin components while atoms traverse trailing atoms after reflecting from the barrier. Note, for instance, the oscillations in only the $\sigma_y$ and $\sigma_z$ components of the front of the cloud as it traverses trailing density which is in the $\ket{x}$ spin state. Meanwhile, atoms at the middle or back of the cloud experience some spin rotation prior to arrival at the barrier. In particular, we see that for large barrier coupling strengths atoms in the back half of the cloud are driven toward $\ket{z}$ prior to arrival at the barrier. Since the pseudomagnetic field of the barrier also points along this direction, the final spin rotations become independent of the external coupling in the strong coupling limit in this experiment, as shown by Fig.~3 of the main text.

\section{Experimental Procedure}
A full description of the experimental procedure can be found in \cite{spierings2021fast} and its corresponding Supplementary Material. The procedures to prepare an atomic wavepacket, implement a pseudomagnetic field inside an optical barrier, and perform spin tomography are identical to the methods there. In addition, the techniques to measure the velocity spread of the atomic cloud, the barrier coupling strength, and the peak energy of the barrier concurrently with the experiment are the same. Here we detail only the differences with those procedures.

\subsection{Varying the barrier coupling strength (i.e.~two-photon Rabi frequency)\label{sec:rabi_frequency}}
In previous work, the Larmor time for tunneling particles was measured by weakly perturbing the potential created by the optical barrier with a pair of Raman beams with a low, $\sim\SI{200}{\Hz}$, two-photon Rabi frequency \cite{spierings2021fast}. The Raman beams are generated through a combination of phase modulation at the $\SI{6.8}{\GHz}$ hyperfine splitting of the ground-state and attenuation of one of the sidebands via an etalon with a bandwidth of $\SI{12}{\GHz}$. The carrier and remaining sideband then couple the $\ket{F=2,m_F=0}$ and $\ket{F=1,m_F=0}$ states. In this experiment, the two-photon Rabi frequency (or ``barrier coupling strength") is varied over a large range, $\sim\SI{100}{\Hz}-\SI{1500}{\Hz}$, by varying the depth of phase modulation. A fast photodiode measures the amplitude of the $\SI{6.8}{\GHz}$ signal and is calibrated via measurements of the two-photon Rabi frequency at the atoms. We use the rotation angles experienced by high-velocity atomic clouds, $\geq\SI{7}{\mm\per\s}$, to calibrate the effective Rabi frequency of the Raman beams, integrated over the Gaussian profile of the barrier. For energies well above the barrier, a semi-classical approach is a good approximation for the time spent traversing the barrier region. Including the spatial profile of the barrier, we use the following equation to define the effective barrier coupling strength $\Omega$:
\begin{equation}
\label{ch:fast:eq:rabi_def}
\theta=\Omega\int_{-\infty}^{\infty} G(y)/\sqrt{v^2-v_\mathrm{b}^2G(y)}dy,
\end{equation}
where $\theta$ is the rotation angle experienced by a high-velocity wavepacket, $G(y)=e^{-2y^2/\sigma^2}$ is the Gaussian profile of the barrier along the ODT direction, $v$ is the velocity of the cloud, and $v_\mathrm{b}$ is the velocity matching the height of the barrier. The above definition introduces the convention that, in the limit of a vanishing barrier height, our calibration for high-velocity atomic clouds corresponds to the time spent in a region of width $\sqrt{\pi/2}\sigma$.

\subsection{Extracting the position-dependent spin profile}

We measure the $S_x$, $S_y$, or $S_z$ components of the transmitted and reflected atomic clouds on a given realization of the experiment via sequential absorption imaging of the $\ket{F = 2, m_F = 0}$ and $\ket{F = 1, m_F = 0}$ populations and a microwave pulse to rotate a given spin component onto the axis read out by the state populations. To extract a two-dimensional spin profile from these absorption images, one would like to compute the difference in the measured atom number in each image pixel-by-pixel. Yet, the population in $\ket{F = 1, m_F = 0}$ moves with respect to the $\ket{F = 2, m_F = 0}$ population in the time between the two absorption images, $\SI{3}{\ms}$. This motion is set by the average incident velocity of the wavepacket and requires a relative shift between the two absorption images, in opposite directions for the transmitted as compared to the reflected cloud. For the slower incident velocities studied this shift is small, about 3 to 4 pixels of the camera, but it becomes significant for faster wavepackets that have energy above the barrier. More exact calculations of the wavepacket motion in the time between the images due to the velocity width of the wavepackets, the velocity filtering due to the barrier, and the motion in the longitudinal potential of the ODT lead to only $\sim0.5$ pixel correction to this shift and are disregarded. 

Once this relative motion is accounted for, the spin profiles are computed by integrating the images along the axis perpendicular to the longitudinal direction of the ODT and calculating the atom number difference pixel-by-pixel in the two absorption images. In other words,
\begin{equation}
    S_i(y)=\frac{N_1(y) - N_2(y)}{N_1(y)+N_2(y)},
\end{equation}
where $y$ is the longitudinal position in the ODT, $N_1$ and $N_2$ are the position-dependent atom numbers in the transmitted or reflected region of the first and second absorption images, and $i=x,y,$ or $z$ depending on the particular realization of the experiment, as described above. This calculation outputs a spin profile that at every position in the transmitted or reflected cloud is normalized to the atom number in that subensemble for the pair of absorption images. Across a dataset we then have an average spin profile, as well as statistical error for that profile.

\section{Interaction-driven spin rotations in the transmitted cloud}

Atoms which transmit the optical barrier in our experiment do not display the intricate spin textures reported here, at least to the accuracy of our current measurements, and the net rotation angles are understood without taking into account interaction-driven rotations \cite{spierings2021fast}. This is a happy accident resulting from the delta-kick cooling procedure that sets the velocity profile of the atomic wavepacket. In this technique, atoms are released from an initial trap providing tight confinement along the longitudinal axis of the ODT. After some expansion, the longitudinally confining potential is pulsed back on in order to reduce the velocity spread of the cloud. The reduction in velocity spread is limited by the Gaussian shape of the pulsed potential, which is more shallow at large displacements from its center than a harmonic potential, and we find that the pulse sequence ‘under kicks’ the atomic cloud. The consequence is that the highest energy components, and thus the most likely to traverse the barrier, are in the front portion of the atomic cloud.

\begin{figure}[t]
    {\centering
    \begin{minipage}[b]{0.49\textwidth}
        \includegraphics[width=\textwidth]{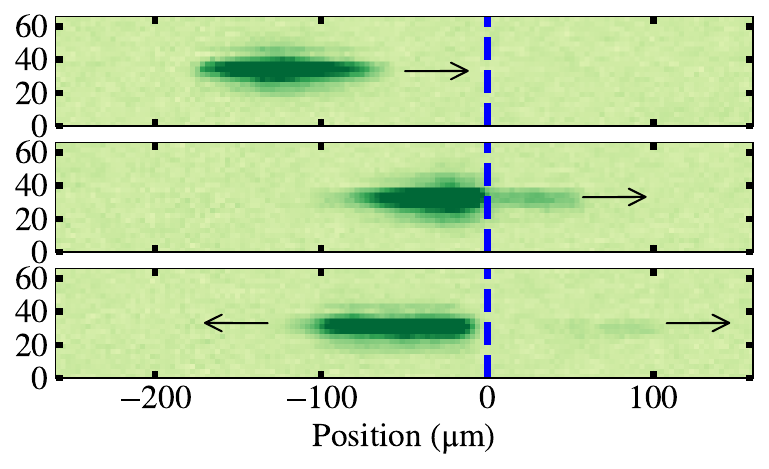}
        % \caption{Caption for Figure 1}
        \begin{picture}(0,0)
        \put(-120,175){ (a)}
        \end{picture}
    \end{minipage}
    \hfill
    \begin{minipage}[b]{0.49\textwidth}
        \includegraphics[width=\textwidth]{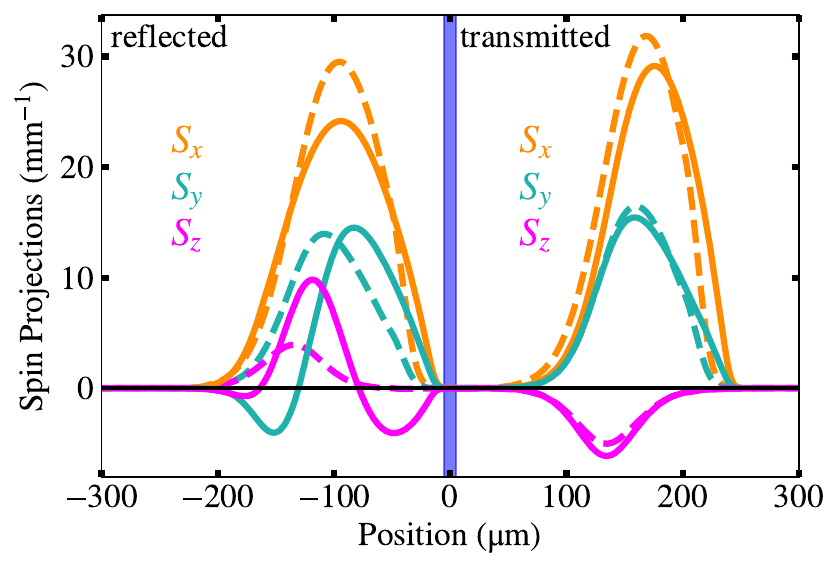}
        % \caption{Caption for Figure 2}
        \begin{picture}(0,0)
        \put(-130,175){ (b)}
        \end{picture}
    \end{minipage}}
    \caption{{\bf Spin profiles in the transmitted cloud are not significantly modified by atomic interactions.} (a) A chronological sequence of absorption images, with time increasing for lower images, demonstrating chirp in the atomic wavepacket. The location of the barrier is indicated by the dashed blue line and the color scheme is intentionally saturated to make the transmitted atoms more visible. Arrows indicate the direction of propagation. (b) A comparison of the spin profiles, $S_x$ (orange), $S_y$ (green), and $S_z$ (magenta) predicted by SE (dashed) and GP (solid) calculations in the reflected and transmitted atomic clouds. As before, the barrier is illustrated by the shaded blue region. The spin profiles are normalized to the particle number in the reflected and transmitted regions.}
    \label{fig:trans_profile}
\end{figure}

This chirp in the velocity profile of the atomic cloud is visible by looking at when the atoms emerge on the far side of the barrier. Figure \ref{fig:trans_profile}(a) depicts three absorption images, before, during, and just after collision with the barrier. The transmitted atoms emerge on the far side well before the reflected cloud has finished interacting with the barrier. As a result, the vast majority of atoms that traverse the barrier do not interact with significant counter-propagating atomic density. Hence, extra spin rotation due to atomic collisions is a much reduced effect for the transmitted atoms. This is well modelled in the simulations throughout this work and in \cite{spierings2021fast}, which mimic the wavepacket preparation of the delta-kick cooling protocol. For example, Fig.~\ref{fig:trans_profile}(b) shows spin profiles in the reflected and transmitted wavepackets predicted by the GP and SE theories. The profiles in the transmitted ensemble show close agreement, whereas the profiles in the reflected cloud display the striking differences discussed in the main text.

\end{document}